\documentclass[final,3p,times]{elsarticle}

\usepackage{amssymb}
\usepackage{amsmath}

\begin{document}

\begin{frontmatter}



\title{Probing CP Violation with Hyperon EDMs at BESIII}


\author{Jianyu Zhang on behalf of the BESIII collaboration} 
\ead{jianyu.zhang@ncbj.gov.pl}

\affiliation{organization={National Centre for Nuclear Research},
            city={Warsaw},
            postcode={02-093}, 
            country={Poland}}

\begin{abstract}
The prevalence of matter over antimatter in the observable universe remains one of the most profound puzzles in modern physics, necessitating sources of charge-parity (CP) symmetry violation beyond those incorporated in the Standard Model. While electric dipole moments (EDMs) of elementary particles serve as sensitive probes for such new physics, the hyperon sector has historically remained a largely unexplored territory. This review synthesizes recent advancements in the search for hyperon EDMs, focusing on the novel modular angular analysis of entangled baryon-antibaryon pairs produced in $J/\psi$ decays at electron-positron colliders. We discuss the theoretical formalism connecting the electric dipole form factor to full angular distributions and highlight the recent milestone achieved by the BESIII experiment, which established an improved upper limit on the $\Lambda$ EDM. This result represents a three-order-of-magnitude improvement over historical measurements. 
\end{abstract}




\end{frontmatter}



\section{Introduction}
\label{sec1}

The question of why the universe is dominated by matter rather than antimatter has driven fundamental research for decades. To generate this observed baryon asymmetry from a symmetric initial state, the Sakharov conditions necessitate the violation of CP symmetry. While CP violation (CPV) has been firmly established in the decays of $K$~\cite{Christenson:1964fg}, $B$~\cite{Belle:2001zzw, BaBar:2001pki}, and $D$~\cite{LHCb:2019hro} mesons, as well as in $\Lambda_b$ baryon decays~\cite{LHCb:2025ray}, these phenomena are accommodated within the Standard Model (SM) via the complex phase of the Cabibbo-Kobayashi-Maskawa (CKM) quark mixing matrix. However, theoretical evaluations indicate that the magnitude of CPV provided by the CKM mechanism, combined with the limits on the QCD vacuum angle $\bar{\theta}$, is insufficient to account for the magnitude of the matter-antimatter asymmetry observed in the universe~\cite{Sakharov:1967dj, Riotto:1998bt}. This discrepancy strongly suggests the existence of new physics (NP) beyond the SM that violates CP symmetry.

In the quest to uncover these new sources of CPV, the measurement of permanent EDMs of elementary particles has emerged as a sensitive and robust probe. A non-zero EDM for a spin-1/2 particle signifies a violation of both Parity (P) and Time-reversal (T) symmetries. Assuming the conservation of CPT symmetry, T violation necessitates CP violation. Unlike CPV in flavor-changing processes, EDMs probe flavor-diagonal CPV, which is highly suppressed in the SM, making any observation of a significant EDM an unambiguous signal of new physics~\cite{Beacham:2019nyx,Chupp:2017rkp}.

Historically, experimental efforts have focused heavily on the neutron and atoms such as $^{199}\text{Hg}$, setting stringent upper limits on the QCD vacuum phase angle, $\bar{\theta}$~\cite{Abel:2020pzs, Graner:2016ses}. However, solely measuring the EDM of first-generation particles (electrons, neutrons) is insufficient to disentangle the potential sources of CPV, such as the quark EDM, the chromo-electric dipole moment (cEDM), and the $\bar{\theta}$ term~\cite{Chupp:2017rkp, Pospelov:2005pr}. A global analysis requires inputs from diverse systems, including the heavy quark or strange quark sectors. Hyperons, which contain one or more strange valence quarks, offer a unique laboratory for this purpose. The strange quark may exhibit specific couplings to BSM fields, potentially enhancing the EDM effect in ways not accessible through neutron measurements~\cite{Chen:2025rab}.

Despite their theoretical importance, hyperon EDM measurements have been hindered by experimental challenges. Direct measurements utilizing spin precession in magnetic fields—standard for long-lived neutrons—are difficult for hyperons due to their extremely short lifetimes ($\sim 10^{-10}$ s). The only prior experimental limit for the $\Lambda$ hyperon, $|d_\Lambda| < 1.5 \times 10^{-16}$ e cm, was established in a fixed-target experiment at Fermilab more than four decades ago~\cite{Pondrom:1981gu}.

 Recent developments at electron-positron colliders, particularly the BESIII experiment, have opened a new avenue for precision hyperon physics~\cite{BESIII:2018cnd, BESIII:2021ypr, BESIII:2022qax, BESIII:2023sgt, BESIII:2023drj, BESIII:2024nif, BESIII:2025jxt, Li:2016tlt}. By exploiting the quantum entanglement of hyperon-antihyperon pairs produced in the decay of the $J/\psi$ resonance, it is possible to extract CP-violating phases from the angular distributions of the decay products~\cite{He:1992ng,He:1993ar,He:2022jjc}. This article details the theoretical framework and experimental realization of this novel method, culminating in the recent BESIII measurement that improved the sensitivity to the $\Lambda$ EDM by three orders of magnitude~\cite{BESIII:2025vxm} and future prospects at the Super Tau-Charm Facility (STCF)~\cite{Fu:2023ose,Achasov:2023gey}.

\section{Theoretical Framework}

\subsection{Effective Lagrangian and CPV Sources}
The electric dipole moment of a fermion is an intrinsic property arising from the displacement of the center of charge from the center of mass along the spin axis. In the non-relativistic limit, the interaction Hamiltonian is given by $H = -d \vec{S} \cdot \vec{E}$, where $\vec{S}$ is the spin and $\vec{E}$ is the electric field. Since $\vec{E}$ is a polar vector (odd under P, even under T) and $\vec{S}$ is an axial vector (even under P, odd under T), a non-zero $d$ requires the violation of both P and T symmetries.

In the context of effective field theory, contributions to the hyperon EDM can be parameterized by a Lagrangian that includes the QCD $\bar{\theta}$ term and dimension-5 and dimension-6 operators representing new physics:
\begin{eqnarray}
    \mathcal{L} = -\frac{\alpha_s}{8\pi}\bar{\theta} \text{Tr}[G^{\mu\nu}\tilde{G}_{\mu\nu}] - \frac{i}{2}\sum_q d_q \bar{q} F_{\mu\nu}\sigma^{\mu\nu}\gamma^5 q \nonumber \\
    - \frac{i}{2}\sum_q \tilde{d}_q \bar{q} G_{\mu\nu}^a T^a \sigma^{\mu\nu}\gamma^5 q
\end{eqnarray}
Here, $d_q$ represents the quark EDM and $\tilde{d}_q$ represents the quark chromo-EDM~\cite{Chupp:2017rkp}. The $\Lambda$ hyperon EDM is sensitive to these parameters, particularly the strange quark contribution ($d_s$), which is largely unconstrained by neutron EDM measurements due to the dominance of $u$ and $d$ quarks in the nucleon wavefunction~\cite{Chen:2025rab}.

\subsection{Form Factors in $J/\psi \rightarrow B\bar{B}$}
In electron-positron annihilation, hyperons are produced in pairs via the process $e^+e^- \rightarrow J/\psi \rightarrow B\bar{B}$. The vertex describing the interaction between the vector meson $J/\psi$ and the baryon pair can be parameterized by Lorentz-invariant form factors. The helicity amplitude for this decay is expressed as:~\cite{He:2022jjc}
\begin{eqnarray}
    \mathcal{M}_{\lambda_1, \lambda_2} = \epsilon_\mu(\lambda_1 - \lambda_2) \bar{u}(\lambda_1, p_1) ( F_V \gamma^\mu + \frac{i}{2m} \sigma^{\mu\nu} q_\nu H_\sigma \nonumber \\
    + \gamma^\mu \gamma^5 F_A + \sigma^{\mu\nu} \gamma^5 q_\nu H_T) v(\lambda_2, p_2)
\end{eqnarray}
In this decomposition, $F_V$ and $H_\sigma$ are the parity-conserving vector and tensor form factors, while $F_A$ and $H_T$ describe parity-violating and CP violating effects, respectively. The term $H_T$ is of paramount importance for EDM searches. It represents a CP-violating vertex mediated by the exchange of a virtual photon between the $J/\psi$ (composed of a $c\bar{c}$ pair) and the hyperon. While $H_T$ is technically a form factor dependent on the squared momentum transfer $q^2 = M_{J/\psi}^2$, in the limit where $q^2 \rightarrow 0$, it corresponds to the static EDM. Assuming the momentum dependence is negligible or treating the form factor as an effective EDM at the production energy, $H_T$ relates to the baryon EDM $d_B$ via the relation:
\begin{eqnarray}
    H_T = \frac{2e}{3M_{J/\psi}^2} g_V d_B
\end{eqnarray}
where $g_V$ is the coupling constant derived from the branching fraction of $J/\psi$ decay to lepton pairs. A non-zero measurement of the $H_T$ serves as a direct indicator of CP violation in the production process.

In addition to CP violation, the production process is sensitive to Parity violation mediated by the weak interaction. The interference between the electromagnetic (virtual photon) and weak (Z-boson) amplitudes introduces a parity-violating contribution characterized by the form factor $F_A$. Within the SM, this term is related to the effective weak mixing angle $\theta_W^{\text{eff}}$. Precise extraction of $F_A$ therefore allows for a determination of $\sin^2\theta_W^{\text{eff}}$ in the charm sector, providing a consistency test of the SM.

The decay $J/\psi \rightarrow \Lambda \bar{\Lambda}$ produces an entangled spin system.  The angular momentum conservation in the reaction ensures that the spin orientation of the $\bar{\Lambda}$ is correlated with that of the $\Lambda$. By reconstructing the subsequent weak decays $\Lambda \rightarrow p \pi^-$ and $\bar{\Lambda} \rightarrow \bar{p} \pi^+$, the self-analyzing nature of these decays allows for the measurement of the hyperon polarization. The differential cross-section is expressed as a modular angular distribution involving the scattering angle of the hyperon ($\theta_\Lambda$) and the decay angles of the proton ($\theta_p, \phi_p$) and antiproton ($\theta_{\bar{p}}, \phi_{\bar{p}}$). The helicity formulation of the full angular distribution is given by~\cite{Fu:2023ose}:
\begin{equation}\label{angdis}
\begin{aligned}
 \frac{d\sigma}{d\Omega}\propto \sum_{\left[\lambda\right]}&R(\lambda_{1},\lambda_{2};\lambda^{\prime}_{1},\lambda^{\prime}_{2})\\
 &D^{*j=1/2}_{\lambda_{1},\lambda_{3}}(\phi_{1},\theta_{1})D^{j=1/2}_{\lambda^{\prime}_{1},\lambda^{\prime}_{3}}(\phi_{1},\theta_{1})\mathcal{H}_{\lambda_{3}}\mathcal{H}^{*}_{\lambda^{\prime}_{3}}\\
 &D^{*j=1/2}_{\lambda_{2},\lambda_{4}}(\phi_{2},\theta_{2})D^{j=1/2}_{\lambda^{\prime}_{2},\lambda^{\prime}_{4}}(\phi_{2},\theta_{2})\overline{\mathcal{H}}_{\lambda_{4}}\overline{\mathcal{H}}^{*}_{\lambda^{\prime}_{4}}\\
 &D^{*j=1/2}_{\lambda_{3},\lambda_{5}}(\phi_{3},\theta_{3})D^{j=1/2}_{\lambda^{\prime}_{3},\lambda_{5}}(\phi_{3},\theta_{3})\mathcal{F}_{\lambda_{5}}\mathcal{F}^{*}_{\lambda_{5}}\\
 &D^{*j=1/2}_{\lambda_{4},\lambda_{6}}(\phi_{4},\theta_{4})D^{j=1/2}_{\lambda^{\prime}_{4},\lambda_{6}}(\phi_{4},\theta_{4})\overline{\mathcal{F}}_{\lambda_{6}}\overline{\mathcal{F}}^{*}_{\lambda_{6}}\\
\end{aligned}
\end{equation}
 Here, $R$ is the spin density matrix encoding the production dynamics (including $H_T$), $D$ are the Wigner rotation functions, and $\mathcal{H}$ represent the weak decay helicity amplitudes.

\section{The BESIII Measurement}

 BESIII has achieved the most precise determination of the $\Lambda$ EDM to date~\cite{BESIII:2025vxm}. The analysis selected events where $\Lambda \rightarrow p \pi^-$ and $\bar{\Lambda} \rightarrow \bar{p} \pi^+$, requiring four charged tracks with zero net charge.  Stringent vertex fitting criteria were applied to identify the displaced decay vertices characteristic of long-lived hyperons, resulting in a signal purity of $99.9\%$ with approximately 3 million candidate events. To prevent experimenter bias, a blinding strategy was employed where the values of sensitive parameters like $H_T$ were obscured during the optimization of the analysis. Systematic uncertainties were rigorously evaluated using data-driven control samples, with the dominant contributions arising from vertex reconstruction efficiency and potential detector resolution effects, though these were found to be small.

Using a data sample of $1.0\times10^{10}$ $J/\psi$ decays, the BESIII Collaboration has set the most stringent limit so far on the $\Lambda$ hyperon electric dipole moment~\cite{BESIII:2025vxm}. The analysis takes advantage of the quantum-entangled $\Lambda\bar{\Lambda}$ pairs produced in $J/\psi\to\Lambda\bar{\Lambda}$, thereby avoiding the difficulties associated with conventional spin-precession methods. A global fit to the joint decay distributions determines the $CP$-odd form factor $H_T$, the parity-violating form factor $F_A$, and the relevant decay parameters, yielding
\begin{eqnarray}
\text{Re}(d_\Lambda) &=& (-3.1 \pm 3.2 \pm 0.5) \times 10^{-19} \, e \, \text{cm}, \nonumber \\
\text{Im}(d_\Lambda) &=& (2.9 \pm 2.6 \pm 0.6) \times 10^{-19} \, e \, \text{cm},
\end{eqnarray}
which are consistent with zero. This leads to the upper bound $|d_\Lambda| < 6.5 \times 10^{-19}\, e\,\mathrm{cm}$ at $95\%$ C.L., improving the previous best limit by about three orders of magnitude.

\begin{figure}
    \centering
    \includegraphics[width=0.45\textwidth]{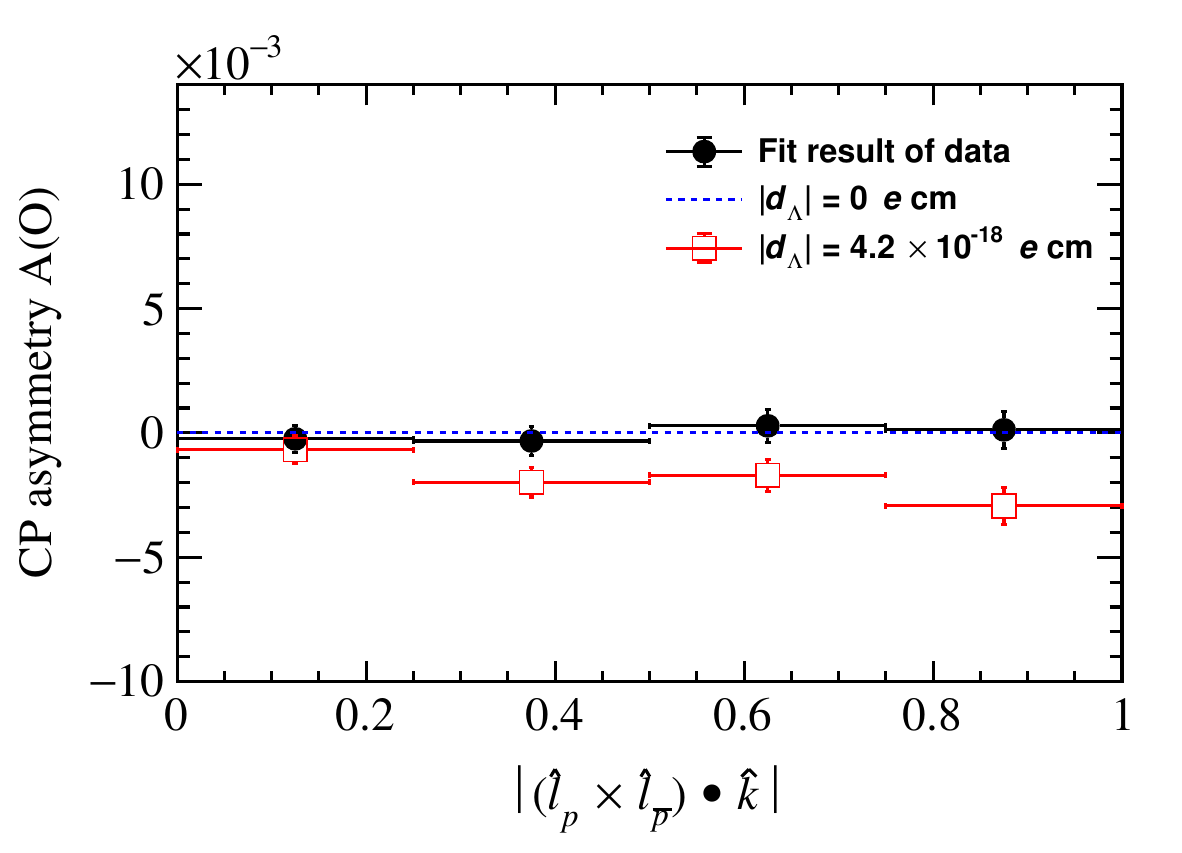}
    \caption{\textbf{Illustration of the $\Lambda$ EDM sensitivity}. The black points show the nominal BESIII result, consistent with zero within uncertainties. The red points correspond to an illustrative nonzero-EDM scenario with $|d_{\Lambda}| = 4.2 \times 10^{-18}\, e\,\mathrm{cm}$, while the blue dashed curve indicates the expectation for vanishing EDM.}
    \label{fig:todd}
\end{figure}

The EDM sensitivity is encoded in $CP$-odd angular correlations of the entangled hyperon pair. A useful quantity is the triple-product observable
\begin{equation}
O\equiv (\hat l_p\times \hat l_{\bar p})\cdot \hat k,
\end{equation}
from which one constructs the asymmetry
\begin{equation}
A(O)=\frac{N_{\text{event}}(O>0)-N_{\text{event}}(O<0)}{N_{\text{event}}(O>0)+N_{\text{event}}(O<0)}.
\end{equation}
This asymmetry is proportional to $\mathrm{Re}(H_T)$ and is shown in Fig.~\ref{fig:todd}. The measured distribution is compatible with the null hypothesis, while a benchmark nonzero EDM case, for example $|d_\Lambda|=4.2\times10^{-18}\,e\,\mathrm{cm}$, would generate a clearly visible deviation.

\begin{figure*}[t]
  \centering
    \includegraphics[width=0.44\linewidth]{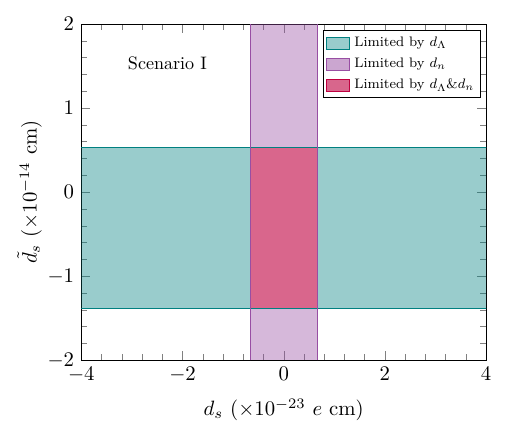}
    \includegraphics[width=0.44\linewidth]{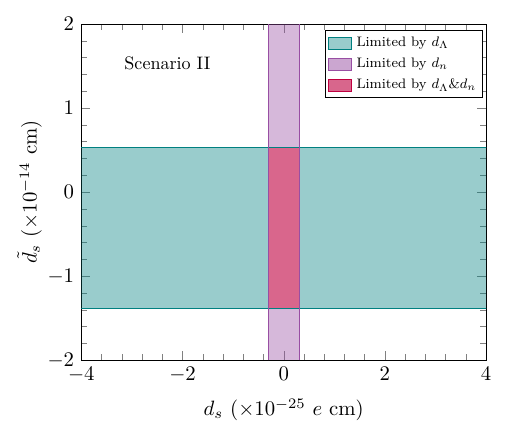}
  \caption{\textbf{Constraints on strange-quark dipole operators from current EDM data.} Panel left shows the allowed region in the scenario $d_s\gg d_u,d_d$, while panel right corresponds to $d_s=d_u=d_d$. The $\Lambda$ EDM mainly improves sensitivity to the strange-quark chromoelectric dipole moment $\tilde d_s$, whereas the neutron EDM more strongly constrains $d_s$.}
  \label{fig:bound}
\end{figure*}

At the theoretical level, the $\Lambda$ EDM measured near the $J/\psi$ scale can be related to quark-level $CP$-violating dipole operators in perturbative QCD~\cite{Chen:2025rab}. Within a collinear factorization treatment, the form factor is expressed as a convolution of hard kernels and hyperon light-cone distribution amplitudes, leading to the expected asymptotic scaling $d_\Lambda(Q)\sim Q^{-4}$. Because the $\Lambda$ is especially sensitive to strange-quark dynamics, its EDM provides information complementary to the neutron EDM. This complementarity is illustrated in Fig.~\ref{fig:bound}: while the neutron EDM sets a stronger constraint on the strange-quark EDM $d_s$, the $\Lambda$ EDM is particularly useful for probing the strange-quark chromoelectric dipole moment $\tilde d_s$. Their combination significantly shrinks the allowed parameter space and leads to the bound
\begin{equation}
    |\tilde d_s| \le 1.4\times10^{-14}\,\mathrm{cm},
\end{equation}
highlighting the role of hyperon measurements in searches for flavor-diagonal $CP$ violation beyond the Standard Model.

\section{Conclusion and outlook}
The search for hyperon EDMs has entered a precision era. By exploiting the entanglement of hyperon pairs at collider experiments, BESIII has improved the limit on the $\Lambda$ EDM by three orders of magnitude, reaching $|d_\Lambda|  < 6.5 \times 10^{-19}$ e cm~\cite{BESIII:2025vxm}. This technique serves as a powerful complement to neutron EDM searches in the global quest to understand the baryon asymmetry of the universe. The methodology established at BESIII is applicable to other hyperons, such as the $\Sigma^+$, $\Xi^-$, and $\Xi^0$~\cite{Fu:2023ose}. Estimates suggest that current BESIII statistics could yield sensitivities of $\sim 10^{-19}$ e cm for these particles as well. The $\Xi$ hyperons are particularly interesting as they contain two strange valence quarks. Looking forward, the proposed STCF is designed to accumulate $3.4 \times 10^{12}$ $J/\psi$ events per year~\cite{Achasov:2023gey}. Projections indicate that STCF could reach EDM sensitivities of $10^{-21} - 10^{-20}$ e cm and measure CP-violating decay asymmetries with a precision of $10^{-5}$~\cite{Fu:2023ose}.

\bibliographystyle{elsarticle-num}
\bibliography{hyperonedm_refs}
\end{document}